\RequirePackage{amsmath}
\RequirePackage{flexisym}
\RequirePackage[superscript,biblabel,nomove]{cite}
\RequirePackage{graphicx}
\documentclass[12pt,aasms4]{article}
\usepackage{amsmath}
\usepackage{amssymb}
\usepackage{setspace}
\def\gsim{\;\rlap{\lower 2.5pt \hbox{$\sim$}}\raise 1.5pt\hbox{$>$}\;}
\def\lsim{\;\rlap{\lower 2.5pt  \hbox{$\sim$}}\raise 1.5pt\hbox{$<$}\;}
\newcommand\beq{\begin{equation}}
\newcommand\eeq{\end{equation}}

\begin{document}

\Large
\centerline{\bf Preliminary Evidence That Protoplanetary}
\centerline{\bf Disks Eject More Mass Than They Retain}

\medskip
\medskip
\normalsize
%\author{}
\centerline{ Amir Siraj$^{1\star}$ \& Abraham Loeb$^1$}
\medskip
\medskip
\medskip
\centerline{\it $^1$Department of Astronomy, Harvard University}
\centerline{\it 60 Garden Street, Cambridge, MA 02138, USA}
\medskip
\medskip
\small
\centerline{$^{\star}$Correspondence to amir.siraj@cfa.harvard.edu}
\normalsize
\vskip 0.2in
\hrule
\vskip 0.2in

\begin{doublespace}
\section*{Abstract}
{\bf }

If interstellar objects originate in protoplanetary disks, they can be used to calibrate the fraction of mass that such disks eject. The discoveries of interstellar objects 1I/`Oumuamua and 2I/Borisov, taken together with rogue planets statistics, allow for the calibration of mass locked in interstellar objects in size range $\sim 10^{4} - 10^{9} \; \mathrm{cm}$. Here, we show that at least $\sim 10\%$ of stellar mass is required to produce the observed population of interstellar objects, with a 95\% confidence interval spanning $\sim 2\% - 50\%$. We call this quantity the Minimum Ejection Fraction (MEF), representing a new constraint on planetary system formation that necessitates an order of magnitude more mass to be processed per star than in the Minimum Mass Solar Nebula (MMSN) model. Future discoveries of interstellar objects with LSST on the Vera C. Rubin Observatory will provide a test of our predictions and improve the statistics.

%\vskip 0.2in

\newpage

\section{Introduction} 
The origin of interstellar objects is an unsolved mystery. Neither exo-Oort clouds nor protoplanetary disks are capable of filling the mass budget necessary to produced the inferred interstellar object population.\cite{2009ApJ...704..733M, 2018ApJ...855L..10D, 2018ApJ...866..131M, 2019AJ....157...86M} More exotic possibilities naturally lead to more implausible mass budgets,\cite{2021JGRE..12606706J, 2021JGRE..12606807D, 2021arXiv210314032S} or issues with survival across interstellar distances and timescales.\cite{2020ApJ...896L...8S, 2020ApJ...899L..23H} Additionally, interstellar objects likely outnumber solar system objects in the Oort cloud.\cite{2020arXiv201114900S} Comets may also be a source of heavy elements in the ISM.\cite{1990ApJ...359..506S} Here, we seek to answer the question: how much mass per star is required to produce interstellar objects? The result will inform future models of interstellar object nurseries and provide constraints on the associated astrophysical processes.

\begin{figure*}[hptb]
%\epsscale{.5}
\includegraphics{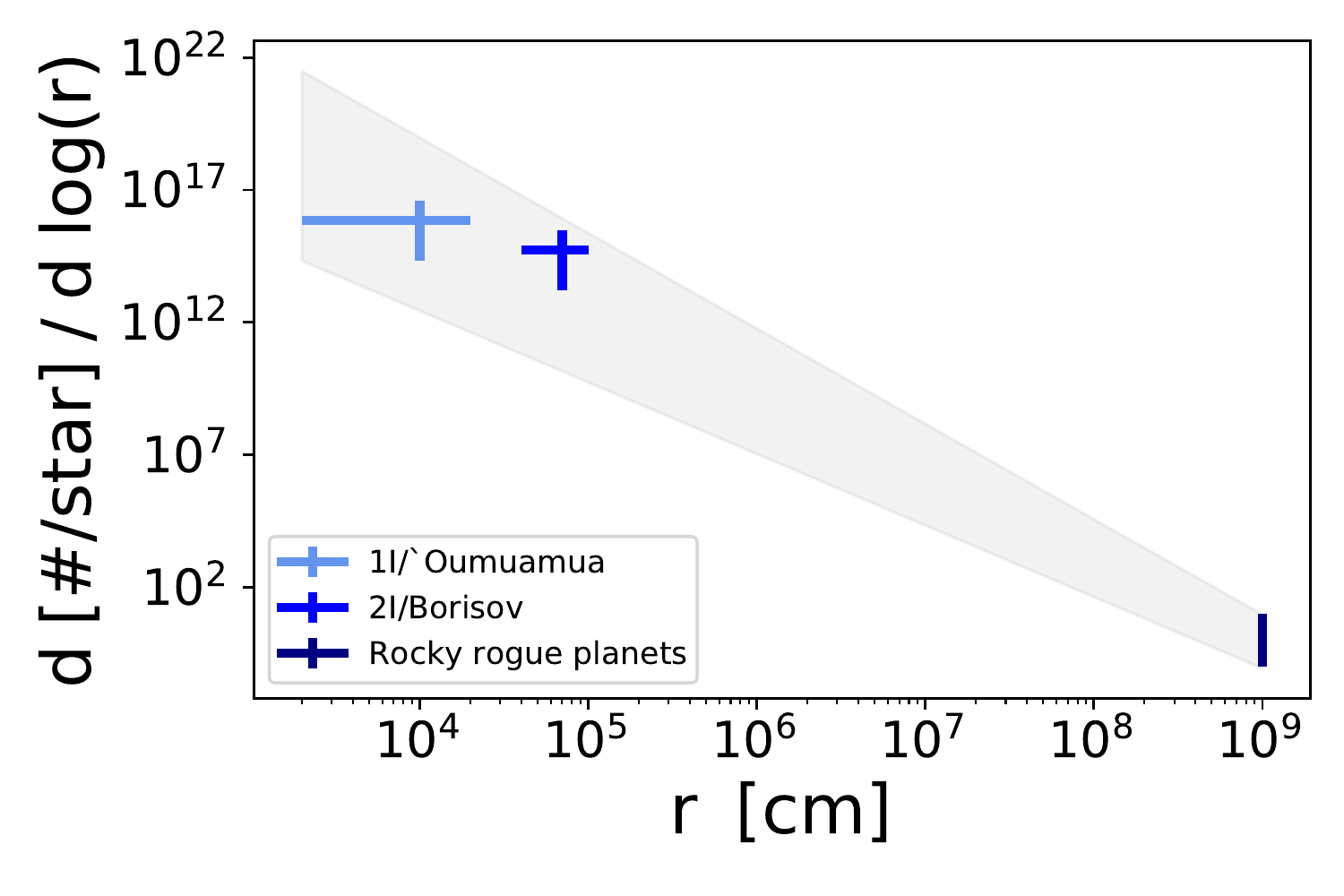}
\caption{\doublespacing `Oumuamua-like objects, Borisov-like objects, and rocky rogue exoplanets in size-abundance parameter space, expressed as number per star per differential unit of log size. The vertical error bars for `Oumuamua and Borisov correspond to the 95\% Poisson uncertainties, while the horizontal error bars for `Oumuamua and Borisov, as well as the vertical error bar for rocky rogue exoplanets, correspond to the discrete ranges discussed in the text.}
\label{fig:sizedist}
\end{figure*}

The first interstellar object, `Oumuamua, was detected in the solar system in 2017.\cite{2017Natur.552..378M} The object's size was estimated to be $20 - 200 \mathrm{\; m}$, based on its lightcurve of reflected sunlight and Spitzer Space Telescope constraints on its infrared emission given its expected surface temperature based on its orbit.\cite{2018AJ....156..261T}. The inferred abundance of similar objects to `Oumuamua is\cite{2018ApJ...855L..10D} $\sim 10^{-1^{+0.98}_{-3}} \; \mathrm{AU^{-3}}$, where the error bars represent the $95\%$ Poisson confidence interval. The discovery of `Oumuamua was followed by that of the second interstellar object, Borisov, in 2019.\cite{2020NatAs...4...53G} The size of Borisov's nucleus was estimated\cite{2020ApJ...888L..23J, 2020arXiv201114900S} to be $0.4 - 1 \mathrm{\; km}$. The inferred abundance of similar objects to Borisov is $\sim 9 \times 10^{-3^{+0.98}_{-3}} \; \mathrm{AU^{-3}}$, \cite{2020arXiv201114900S} where the error bars represent the $95\%$ Poisson confidence interval. Finally, rocky rogue planets with a radius of $\sim 10^{9} \mathrm{\; cm}$, roughly twice the Earth's radius, may number $\sim 5 - 10$ per star,\cite{2017Natur.548..183M} but other estimates range as low as $\sim 2$ per star.\cite{2020AJ....160..123J} Figure \ref{fig:sizedist} illustrates the three classes of interstellar objects considered here in size-abundance parameter space. 

\section{Monte Carlo Simulation} 

Given the sizes and abundances of known interstellar objects, how much mass is required, per star, to produce such a population? We address this question with a Monte Carlo simulation. For each run of the simulation, we select a size and a population abundance for `Oumuamua, Borisov, and rocky rogue planets. The sizes for the `Oumuamua and Borisov are drawn from the triangular distributions spanning $20 - 200 \mathrm{\; m}$ and $0.4 - 1 \mathrm{\; km}$, respectively, with peaks at $100 \mathrm{\; m}$ and $0.7 \mathrm{\; km}$. Rocky rogue planets are taken to have a mean radius $\sim 10^9 \mathrm{\; cm}$. The abundances for `Oumuamua and Borisov are drawn from the Poisson distributions for sample size of one with central values of $10^{-1} \mathrm{\; AU}$ and $9 \times 10^{-3} \mathrm{\; AU}$, respectively. The abundance of rocky rogue planets are drawn from a uniform distribution spanning the range, $1 - 10$ per star. Given the three size-abundance tuples, each run of the simulation fits a linear least square to the log-log transformed values, resulting in a power-law size distribution fit of the form $dN/dR \sim k R^{-q}$ with normalization $k$ and slope $q$. We then derive the total interstellar object mass density for objects larger than `Oumuamua, $\rho_I \sim \int_{R_{O}}^{R_{P}} (dN/dR) \cdot (4\pi \rho_{obj}/ 3) R^3 dR$, where $\rho_{obj} \sim 3 \mathrm{\; g \; cm^{-3}}$ is rock density, $R_O$ is the radius for `Oumuamua drawn from the aforementioned distribution, and $R_P \sim 10^9$ is the fiducial rocky rogue planet radius. We repeat this routine $10^6$ times, thereby constructing a distribution of $\rho_I$ corresponding to the uncertainties in `Oumuamua-like, Borisov-like, and rocky rogue planet populations. Our model applies if `Oumuamua was not primarily composed of hydrogen or helium. A third possible interstellar object, CNEOS-2014-01-08, is fully consistent with the mass budget derived here.\cite{2019arXiv190407224S, 2019arXiv190603270S}

\section{Results \& Discussion} 

The mass fraction of interstellar objects per star is, $M_{ISO}/M_{\star} = \rho_I / \rho_{\star}$, where $\rho_{\star} \sim 4 \times 10^{-2} \; \mathrm{M_{\odot} \; pc^{-3}}$ is the local stellar density\cite{2017MNRAS.470.1360B}. We find that $\rho_I \sim 5 \times 10^2 \; M_{\oplus} \mathrm{\; per\;} M_{\odot}$ with a 95\% confidence interval of $10^2  - 2 \times 10^3 \; M_{\oplus} \mathrm{\; per\;} M_{\odot}$. How does the mass locked in interstellar objects compare to the mass locked in their possible sources? Let us first consider the solar system's Oort cloud. The Oort cloud was estimated to carry a mass of $1.9 M_{\oplus}$,\cite{1996ASPC..107..265W} with more recent simulations suggesting $0.5 - 1.4 M_{\oplus}$.\cite{2000Icar..145..580F, 2008A&A...492..251B} Since $25 - 65\%$ Oort cloud is ejected over time,\cite{2018MNRAS.473.5432H} the range for initial Oort cloud mass encompassing all of the estimates is $0.7 - 5.4 \; M_{\oplus}$. The upper end of this range is only $\sim 1\%$ of the central value for the total mass of interstellar objects. As a result, Oort clouds of stars are implausible sources of interstellar objects. Let us next consider protoplanetary disks. The Sun's protoplanetary disk was estimated to contain $12 - 65 \; M_{\oplus}$ of solids.\cite{2013ApJ...768...45N, 2014ApJ...792..127R, 2017AJ....153..153D} 
Around other stars, only $3 - 4\%$ of observed protoplanetary disks have at least $\sim 10^2 M_{\oplus}$ in solids, although the true proportion may be greater.\cite{2020ARA&A..58..483A} Measurements of protoplanetary disk masses, to date, are inconsistent with the inferred mass of the interstellar object population. We note that considering interstellar objects smaller than `Oumuamua would deepen this tension.

\begin{figure*}[hptb]
%\epsscale{.5}
\includegraphics{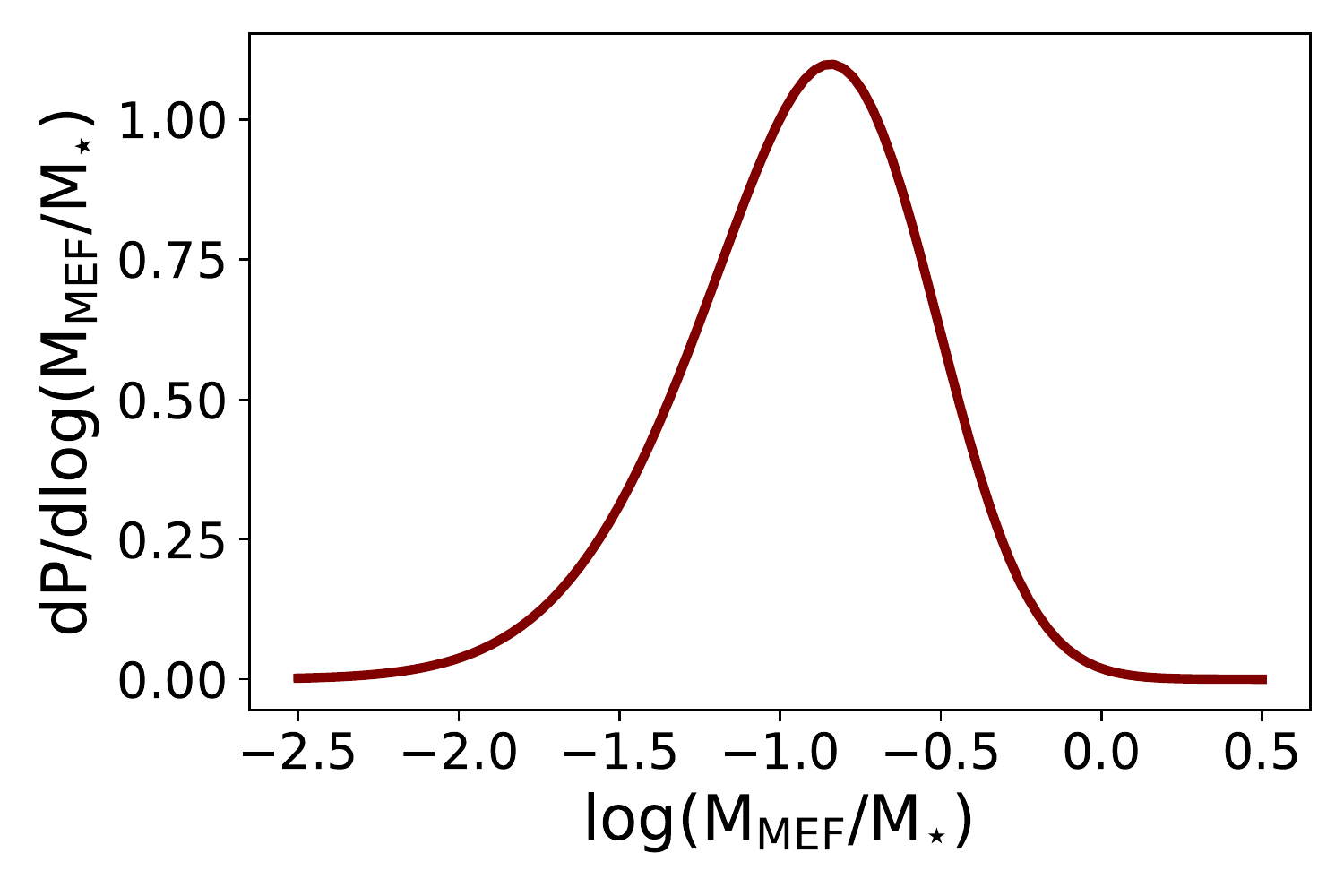}
\caption{\doublespacing Normalized probability distribution for $(M_{MEF} / M_{\star})$, the mass fraction per star necessary to generate the observed of population of interstellar objects larger than `Oumuamua.}
\label{fig:MEF}
\end{figure*}

The Minimum Mass Solar Nebula (MMSN) model requires $\sim 1\%$ of the Sun's mass to form the planets. \cite{1977MNRAS.180...57W, 1981PThPS..70...35H, 2007ApJ...671..878D, 2009ApJ...698..606C} How much mass per star is necessary to form the population of interstellar objects larger than `Oumuamua, and how does this compare to the MMSN? Given the Sun's metallicity of 1.4\%,\cite{2009ARA&A..47..481A} and since `Oumuamua-like objects, Borisov-like objects, and rocky rogue planets are all primarily composed of elements heavier than hydrogen and helium by mass: the total mass budget necessary to produce this population of interstellar obejcts, which we dub the Minimum Ejection Fraction (MEF), is related to the mass budget of interstellar objects as follows, $M_{MEF} = (M_{ISO} / 1.4\%)$. Given the Monte Carlo simulation defined above, we find that the central value to be $M_{MEF} \sim 0.1 M_{\star}$, with a 95\% confidence interval of $0.02 M_{\star} - 0.5 M_{\star}$. Figure \ref{fig:MEF} shows the overall probability distribution for $M_{MEF} / M_{\star}$.

%\newpage 

The primary implication of these results is that the amount of mass needed to form interstellar objects larger than `Oumuamua is a substantial fraction of host star mass, between 2\% and 50\%. This provides a new constraint on planetary system formation, since it requires an order of magnitude more mass than the MMSN,\cite{1977MNRAS.180...57W, 1981PThPS..70...35H, 2007ApJ...671..878D, 2009ApJ...698..606C} which is comparable to previous lower bounds on the interstellar object mass budget.\cite{2020arXiv201114900S} These results suggest a highly efficient route for converting protostellar matter into $\gtrsim 0.1 \mathrm{\; km}$ planetesimals and for ejecting them from their parent stars, and changes the paradigm regarding observational constraints on the planetary system formation process.

Additionally, these results imply that the ejected mass from stars exceeds the retained mass. Neither the mass budget of the solar system's protoplanetary disk\cite{2013ApJ...768...45N, 2014ApJ...792..127R, 2017AJ....153..153D} nor the observed protoplanetary or debris disks around other stars could provide sufficient material for the formation of interstellar objects.\cite{2020ARA&A..58..483A} Large planetesimals must be ejected by dynamical processes, and such processes must provide kicks strong enough to unbind them from their parent stars. Stellar binaries are likely to eject the inner region of protoplanetary disks\cite{2018ApJ...866..131M, 2018MNRAS.478L..49J} out to approximately twice their semimajor axis for disks larger than the binary separation \cite{2011ASL.....4..181C, 2020ApJ...900...43T}. Giant planets could create a ring/gap substructure in their protoplanetary disks by ejecting planetesimals from their vicinity.\cite{2019ApJ...884L..22R} The origins of interstellar objects can be inferred through their velocity distribution, once a sufficient number of them have been detected. \cite{2020ApJ...903L..20S} 

\end{doublespace}

\vskip 0.45in
\hrule
\vskip 0.15in

\small
\noindent

%\bibliographystyle{mnras}
%\bibliography{bib} %{}

%\bibliographystyle{naturemag}
%\citestyle{nature}
%\bibliography{bib}

%\printbibliography
\bibliography{p.bib}

%\end{thebibliography}
%\vskip 1in

\end{document}